\documentclass[twocolumn,showpacs,preprintnumbers,floatfix,prl]{revtex4}

\usepackage{graphicx}
\usepackage{subfigure}
\usepackage[]{amssymb}
\usepackage[]{amsmath}

\begin{document}

\title{How long will a bubble be?}

\author{T. Gilet, T. Scheller, E. Reyssat$^{\dag}$, N. Vandewalle, and S. Dorbolo}

\affiliation{GRASP, Physics Department, University of Li\`ege, B-4000 Li\`ege, Belgium} \affiliation{$^{\dag}$
PMMH, UMR 7636 du CNRS, ESPCI, 10 Rue Vauquelin, 75005 Paris, France}

\begin{abstract}
A soap bubble is a metastable object that eventually breaks. Indeed, the soapy water film thins until rupture,
due to drainage and evaporation. In our experimental investigations, floating bubbles at the surface of a liquid
bath have been considered. Their lifetime has been measured and reported with respect to their radius. Large
bubbles last longer than small ones. Moreover, small bubbles have more predictable lifetimes than large ones. We
propose a general equation for that lifetime, based on the lubrication theory. The evaporation is shown to be an
essential process which determines the bubble lifetime.
\end{abstract}
\maketitle

A child succeeds in blowing a soapy bubble, spherical, light, fragile. The bubble flies through the air, avoids
some mortal hydrophobic obstacles and eventually dies after a frontal collision with a wild red poppy. How long
would it have lived ? A second question should also be addressed: is the bubble lifetime related to its size
when prevented from any accident ? Some indications can be found in the literature. The lifetime of large
bubbles made of PDMS oil with a large viscosity (1000Pa$\cdot$s) have been studied in \cite{Debregeas:1998}.
This liquid had been chosen to avoid dust contamination and evaporation. The bubble lifetime has been shown to
be related to the thinning of the film at the top of the bubble. The thickness is found to decrease according to
an exponential law until a critical value for which the film breaks. This allows to define a lifetime $\tau$ for
the bubble. In the case of PDMS bubbles, the lifetime has been found to be inversely proportional to the bubble
radius. Although the thinning behavior has also been observed in surfactant-water systems \cite{Angarska:2001},
flows in soapy films are very different from the ones in the PDMS system. Firstly, surfactant molecules rigidify
the interface: The flow cannot be described as a plug flow. Secondly, the evaporation cannot be neglected.
Indeed, let us remind that when a soap film is prevented from any evaporation and nucleation by dusts, the film
is metastable and the bubble may last as long as the experimentalist's patience. On the other hand, physical
processes as drainage, nucleation of holes by dusts and evaporation reduce the lifetime of the bubble. Are those
processes influenced by the size of the bubble? Some experimental works can be found about the lifetime of
n-butanol and n-nonanol bubbles \cite{Warszynski:1996,Jachimska:1998}. In these papers, the motion of the
bubbles through the liquid is shown to modify their lifetime because the surfactant molecules located at the
interfaces are redistributed during the travel.

\begin{figure}[htbp]
\subfigure[]{\label{fig:0005mlphoto2}
\includegraphics[width=\columnwidth]{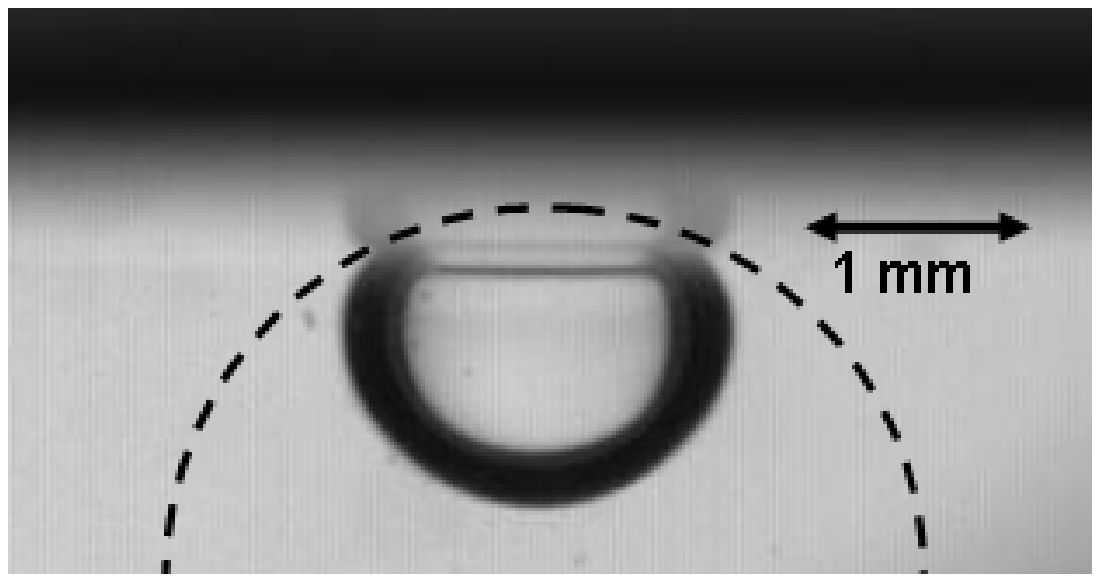}}
\subfigure[]{\label{fig:80mlphoto}
\includegraphics[width=\columnwidth]{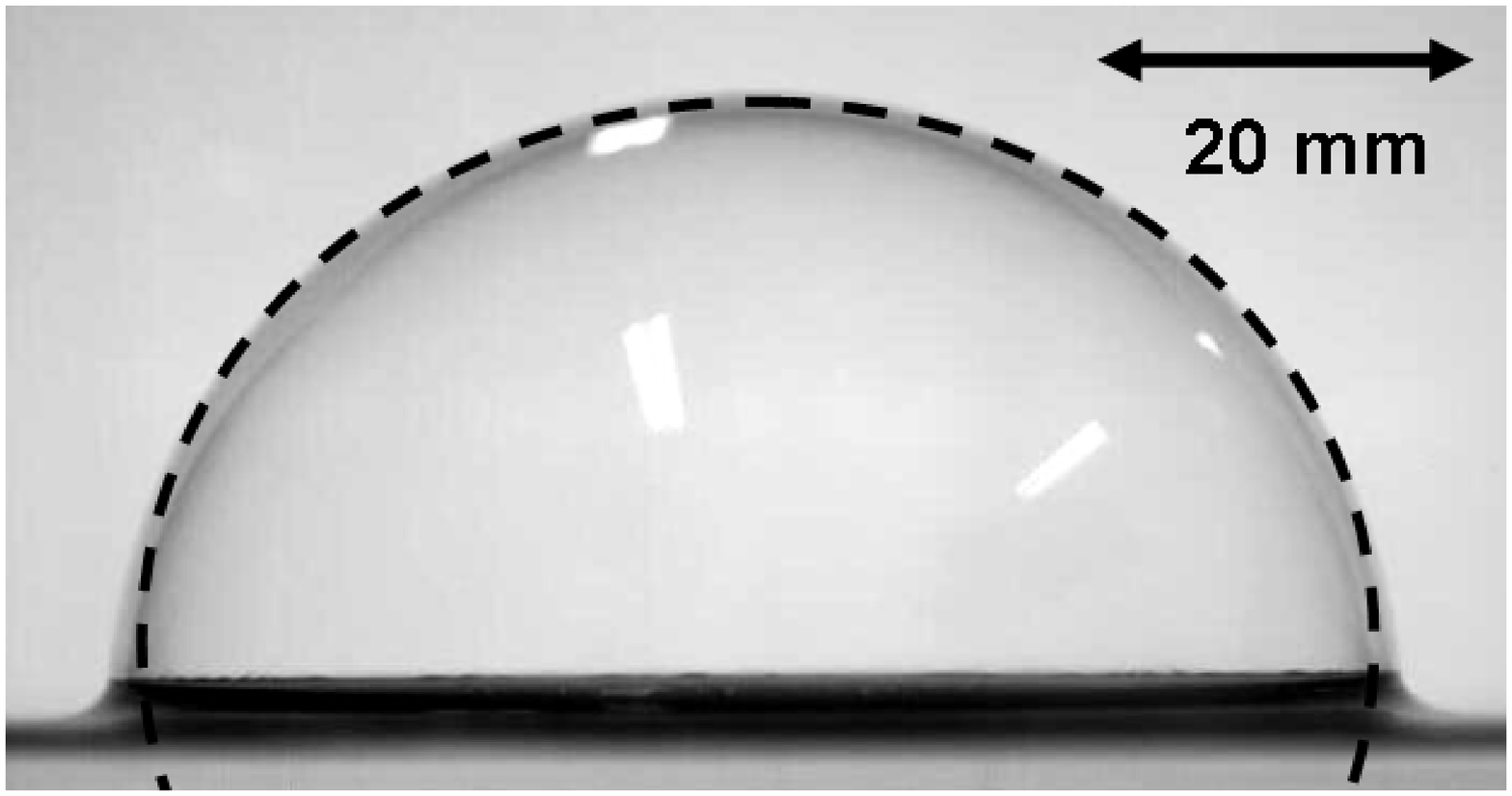}}
\caption{\label{fig:Photos} Pictures of bubbles at liquid-air interfaces. The upper part of the bubble (above
the bath interface) can be fitted by a sphere of radius $R$ (Here approximately drawn). (a) Bubble with
$R=1.67$mm (0.005mL in volume); the air bubble remains below the liquid-air interface. (b) Bubble with
$R=31.6$mm (80mL in volume); the bubble shape is an emerged hemisphere.}
\end{figure}

In this letter, the lifetime of a single bubble of soapy water is investigated. The original part of this work
is that evaporation plays here a determinant role in the bubble lifetime. In order to stabilize the bubble,
surfactant Triton X-100 is added to water with a concentration of 2mMol/L. This corresponds to approximately 10
times the critical micellar concentration. Indeed, the lifetime of the bubble increases with the surfactant
concentration, and even saturates for large concentrations \cite{Jachimska:1998,Warszynski:1996}. Therefore,
using a liquid with a concentration of surfactant 10 times above the CMC prevents from any effects due to the
depletion of surfactant molecules. Moreover, interfaces can be considered as nearly rigid ones and the plug-flow
component is negligible. That surfactant molecule is non-ionic and is used as a standard in many
physico-chemical experiments.

Contrary to the previous works \cite{Jachimska:1998,Warszynski:1996,Debregeas:1998}, bubbles are created at the
surface of a water bath by using a syringe filled with air. In so doing, the distribution of the surfactant
molecules along the surface of the bubble is uniform and relatively homogenous. The volume of the bubble is
tuned by using different syringes, from 2$\mu$L to 40mL. The relative error made on the volume is estimated to
be less than 5\%, mainly due to the graduation reading. The atmospheric conditions are constant during the whole
experiment. In particular, the temperature is 22$^o$C and the relative humidity $\zeta=0.54$. In order to avoid
dust contamination and hazardous variations of the evaporation rate, the system is set up in a constant laminar
flow of purified air with a velocity about 0.01m/s (Captair Flow - Erlab). Moreover, that ensures the
atmospheric perturbation (wind) to be constant and the same for all our experiments. A digital camera records
the bubble from the side (Fig.\ref{fig:Photos}). The relative error on the radius measurement is less than 2\%
(due to the resolution of the camera). Moreover, an average of the geometrical properties is made over three
different bubbles with the same volume.

\begin{figure}[htbp]
\includegraphics[width=\columnwidth] {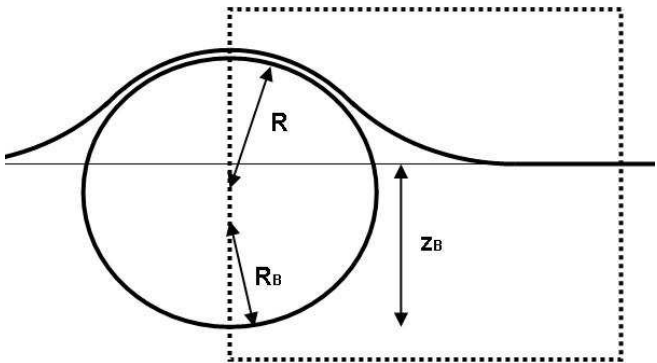}
\caption{\label{fig:Equilibrium} Shape of a bubble near a liquid-air interface. Radii of curvature at the bottom
$R_B$ and at the top $R$ are related to the bubble depth $z_B$ according to Eq.(\ref{eq:Equilibrium}). The
dotted curve corresponds to the contour along which pressure variations are integrated.}
\end{figure}

The shape of a bubble (or a droplet) near a liquid-air interface is the result of a competition between reduced
gravity (gravity minus buoyancy effects) and surface tension effects \cite{Plateau:1873}. When gravity is
negligible compared to surface forces, the bubble remains spherical and fully submerged below the perfectly
planar surface of the bath (Fig.\ref{fig:0005mlphoto2}). This configuration minimizes the surface energy of the
system. When gravity is much more important than interfacial tension, the bubble is nearly hemispherical and
fully emerged (Fig.\ref{fig:80mlphoto}). Between those regimes, the interface of the bath is slightly deformed
by the bubble. The upper part of the bubble (above the meniscus) is approximated as a spherical cap
\cite{Plateau:1873}. This is merely due to a nearly perfect balance in hydrostatic pressure inside and outside
the bubble (the weight of the film being negligible). The radius of this spherical cap is denoted $R$; it is the
relevant parameter for the determination of the lifetime. Since small bubbles are fully immersed, the upper part
is not always visible (as in Fig.\ref{fig:0005mlphoto2}). The radius is then estimated by the following pressure
balance \cite{Boucher:1980,Jones:1978}:
\begin{equation} \label{eq:Equilibrium}
4 \sigma/R = 2 \sigma/R_B + \rho_w g z_B
\end{equation}
where $R_B$ is the radius of curvature at the bottom of the bubble, $\rho_w$ the water density, $z_B$ the depth
of the bottom and $\sigma$ the interfacial tension. This balance is deduced from an integration of pressure
variations along the dotted contour in Fig.\ref{fig:Equilibrium}.

\begin{figure}[htbp]
\includegraphics[width=\columnwidth] {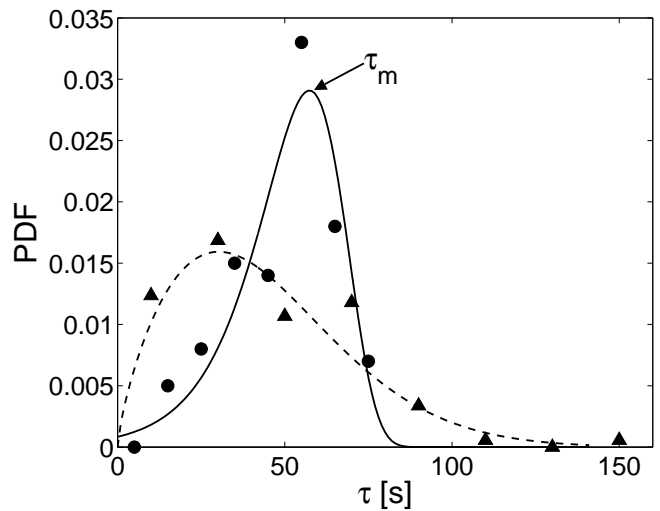}
\caption{\label{fig:Statistics} PDF of the lifetime $\tau$ for two given bubble volumes: ($\bullet$) 6.07mm -
($\blacktriangle$) 25.1mm. Curves are guides for the eyes. The solid line is an extreme value distribution with
a mean $\tau_m = 57$s and a standard deviation $\sigma = 12.6$s; the dashed line is a Weibull distribution with
a scale parameter of 49.5s and an exponent of 1.73.}
\end{figure}

The statistical distribution of the lifetimes $\tau$ is shown for two given bubble radii: $R=6.07$mm (0.5mL in
volume) and $R=25.1$mm (40mL in volume). A hundred bubbles have been studied for each size. In
Fig.\ref{fig:Statistics}, the Probability Density Function (PDF) of the lifetime $\tau$ has been plotted. For
the 6.07mm bubble, the PDF is a right-shifted peak that is well-fitted by an extreme value distribution. Such a
result has also been found in \cite{Jachimska:1998,Warszynski:1996}. It suggests that the lifetime related to
the peak is the determinist maximum lifetime $\tau_m$ of the bubble. Lower lifetimes are due to random
accidents. These accidents cannot occur after the determinist lifetime. For the 25.1mm bubble, the distribution
seems to behave like a broad Weibull distribution: no determinist trend can be emphasized. Experimentally, the
lifetime of small bubbles ($R<10$mm) has been observed to roughly obey a right-shifted peak distribution, while
the lifetime of large bubbles ($R>10$mm) follows a broad and more or less uniform distribution.

\begin{figure}[htbp]
\includegraphics[width=\columnwidth] {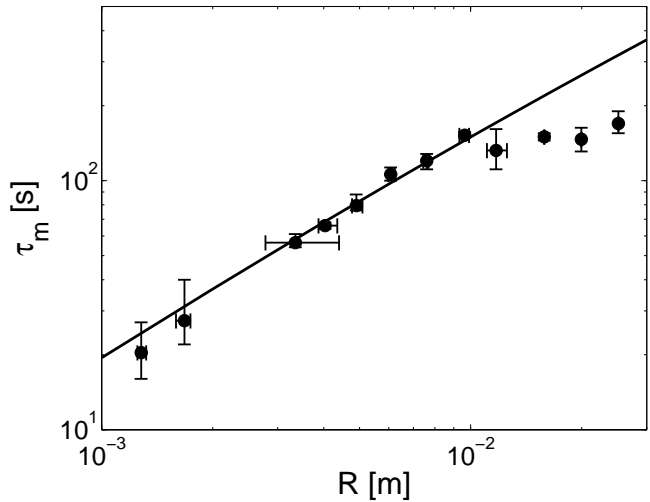}
\caption{\label{fig:Lifetime} Maximum lifetime $\tau_m$ of the bubbles as a function of the radius $R$ of the
upper spherical cap. The line is related to the law described by Eq.(\ref{eq:Lifetime}).}
\end{figure}

For each bubble size, about 10 measurements are made. When the lifetime $\tau$ is predictable and obeys the
first kind PDF, it makes sense to only consider the mean of the 5 greatest observed lifetimes in order to remove
the premature accidental deaths. This averaged maximum lifetime $\tau_m$ is presented with respect to the
spherical cap radius in Fig.\ref{fig:Lifetime}. For bubbles with a radius $R \lesssim 10$mm, $\tau_m$ is roughly
proportional to $R$. For larger bubbles, $\tau_m$ deviates from the linear law and the lifetime distribution
becomes more spread. This lifetime variation with the bubble size is completely different from the one obtained
with PDMS bubbles \cite{Debregeas:1998} (the lifetime was inversely proportional to the radius).

As mentioned in the introduction, the lifetime is related to the thinning of the film located at the top of the
bubble. The reasons for that decrease of the film thickness $h$ are \textit{(i)} the drainage by gravity and
\textit{(ii)} the evaporation. Due to the high surfactant concentration, we can reasonably assume that the
interfaces of the bubble are rigid ones. The situation is again different from the PDMS bubble
\cite{Debregeas:1998} for which a plug flow is considered since the interface is not rigidified by surfactant
molecules. At the top of the bubble, they are well approximated by a spherical cap of radius $R \pm h/2$.

The water film between both surfactant layers is thin enough to suppose that the flow obeys Poiseuille law. The
tangential velocity $U$ of this flow, averaged over the film thickness $h$, is given by $U = - (h^2
\partial_{\theta} p) / (12 \mu_w R)$ where $\theta$ is the azimuthal coordinate (starting from
0 at the top), $\mu_w$ the dynamical viscosity of the water, and $p$ the pressure inside the film. The pressure
gradient in the tangential direction is mainly due to gravity: $\partial_{\theta} p = - \rho_w g R \sin \theta$,
where $\rho_w$ is the density of water.

The drainage is particularly slow when the film is very thin. In these conditions, the evaporation process
becomes dominant compared to the drainage. Maxwell \cite{Maxwell:1890} and Langmuir \cite{Langmuir:1918} have
independently proposed models of evaporation based on the diffusion of vapor molecules into the air. The radial
mass flux of vapor $J$ is given by Fick's law as a function of the mass fraction of vapor $m_v = \rho_v /
\rho_a$, where $\rho_v$ is the density of vapor and $\rho_a$ the density of air: $J = -\rho_a D_v dm_v / dr$. In
those models, the diffusion is supposed to be steady and the convection in the air is neglected. Practically,
the first hypothesis is supported by the fact that the diffusion coefficient of water vapor in air is about $D_v
= 2.6 \times 10^{-5}$m$^2$/s. The relaxation time is about 4s for a centimetric bubble. The second hypothesis is
roughly verified since in the experimental setup, the air velocity $V_a$ is very small near the bubble ($V_a
\sim 0.01$m/s). Ranz and Marshall \cite{Ranz:1952} have proposed a model to take convection into account.
According to them, the flux is given by $J R =(1 + 0.42 Re^{1/2} Sc^{1/3}) \rho_a D_v [ m_v(R) - m_v(+\infty)
]$, where $Re=V_a R / \nu_a$ is the Reynolds number of the convection flow, and $Sc=D_v/\nu_a$ is the Schmidt
number. Since at the interface, the vapor pressure $p_v$ is equal to the saturation vapor pressure $p_v^{sat}$,
the difference in mass fraction can be expressed as a function of thermodynamical properties of vapor in air:
$m_v(R) - m_v(+\infty) = ( 1 - \zeta) (M_v p_v^{sat})/(M_a p_a)$ where $M_v$ and $M_a$ are respectively the
molar mass of vapor and air, $p_a$ is the partial pressure of the air and $\zeta$ is the relative humidity. From
these considerations, the evaporation rate can be inferred:
\begin{equation}
\begin{array}{l}
\frac{J(R)}{\rho_w} = \frac{k}{R} \mbox{ , with } \\
k = \frac{\rho_a}{\rho_w} D_v \frac{M_v}{M_a} \frac{p_v^{sat}}{p_a} (1-\zeta) (1+0.42 Re^{1/2} Sc^{1/3})
\end{array}
\end{equation}
Since $p_v^{sat} \simeq 0.031 p_a$ at 25°C and $\zeta = 0.54$, the evaporation coefficient is about $k \simeq
2.8 \times 10^{-10}$m$^2$/s. Due to convection, this coefficient is higher (about $0.5 \times 10^{-10}$m$^2$/s)
and it slightly depends on the Reynolds number (and thus on bubble radius $R$).

The evolution of the film thickness is found by using the continuity equation, and the following lubrication
equation is obtained:
\begin{equation} \label{eq:LubrifAng}
\partial_t h + \frac{1}{\sin \theta} \partial_{\theta} \biggl( U \frac{h}{R} \sin \theta \biggr) + \frac{k}{R}= 0
\end{equation}
As observed and explained in \cite{Couder:2005}, the angular dependence is weak, in particular for $\theta < \pi/4$: the film thickness is
roughly constant in space over the upper part of the bubble. Moreover, the location of the minimum in thickness is obviously the top of the
bubble. Therefore, we can linearize Eq.(\ref{eq:LubrifAng}) for positions near to $\theta = 0$, leading to:
\begin{equation}
\frac{d h}{dt} + \frac{\rho_w g}{6 \mu_w R} h^3 + \frac{k}{R} = 0
\end{equation}
The maximum lifetime of a bubble is then given by
\begin{equation} \label{eq:LifetimeFull}
\tau_m =\frac{R}{k} \int_{h_c}^{h_0} \frac{dh}{1 + \frac{h^3}{\beta^3}}
\end{equation}
where $h_0$ is the initial film thickness, $h_c$ the film thickness at rupture \cite{Angarska:2004} and $\beta^3
= 6 \mu_w k / \rho_w g$. The $\beta$ parameter is interpreted as the length scale at which evaporation and
drainage are equally efficient. For $h \ll \beta$, the film thins by evaporation much more than by drainage,
while it is the opposite for $h \gg \beta$.

On a practical point of view, $h_c \approx 10^{-7}$m while $\beta \approx 6 \times 10^{-6}$m. Therefore, $h_c$
can be replaced by 0 in Eq.(\ref{eq:LifetimeFull}). In a similar way, when the initial thickness $h_0$ is much
higher than $\beta$, $h_0$ can be replaced by $+\infty$ without changing significantly the resulting lifetime
$\tau_m$. Therefore, one has
\begin{equation} \label{eq:Lifetime}
\tau_m \simeq  \pi R \sqrt[3]{\frac{16 \mu_w}{27 \sqrt{3} \rho_w g k^2}}
\end{equation}
According to this expression, the lifetime $\tau_m$ is roughly proportional to the radius of the spherical cap
$R$ ($k$ slightly depends on $R$). This law is plotted in Fig.\ref{fig:Lifetime}; it is in good agreement with
the experiments, at least for small bubble sizes. Note that no fitting parameter is needed.

While $h_0 \gg \beta$, the lifetime is predictable since it does not depend on initial and final thickness
values. The deviation to the theoretical law reflects the manner that we produce the bubble. To obtain a large
bubble, some air is blown inside the liquid shell. The inflation thins the film; the initial film thickness
$h_0$ (after blowing) is thus lowered. When bubbles are large enough, $h_0 \leq \beta$ and it cannot be replaced
by infinity in the lifetime computation. The lifetime now depends on $h_0$, and is smaller than its asymptotic
value. Since $h_0$ is not controlled at all, $\tau$ appears to be not determinist, as seen in the statistical
analysis. The change in the PDF shape occurs for the same bubble size than the deviation from the linear scaling
for $\tau_m$.

In conclusion, an experimental work has been made in order to assess about the lifetime $\tau$ of single soap
bubbles created at the surface of a water bath. The maximum observed lifetime $\tau_m$ is shown to be roughly
proportional to this radius for small bubbles. According to a statistical analysis, this maximum seems to be
determinist and reproductible. Contrary to previous works, the evaporation seems to be the key physical process
that determines the lifetime. The bubble is geometrically characterized by the radius $R$ of the upper spherical
cap that is formed when the bubble rises the surface of the bath. An analytical model, based on drainage and
evaporation of the water film under a small convection flow, has been proposed in order to explain this
quasi-linear scaling. When the initial film thickness is large enough (typically for small bubbles), the
lifetime does not depend anymore on initial and final conditions: it is predictable. When bubbles are inflated
too much, the initial film thickness becomes small enough to infer on the lifetime by reducing it. Since this
thickness is not controlled, the lifetime looks unpredictable. When several bubbles are considered together (as
in foams \cite{Caps:2005}), Plateau borders probably play a significant role in the drainage process and the
variation of the lifetime with the size is expected to be different.

TG and SD thank FNRS-FRIA for financial support. Exchanges between laboratories have been financially helped by
the COST action P21. Part of this work has been supported by Colgate-Palmolive. Prof. K. Malysa (Krakow), Dr. H.
Caps and D. Terwagne are acknowledged for fruitful discussions.


\end{document}